\documentclass[preprints,article,submit,pdftex,moreauthors]{./mdpi} 

\usepackage{todonotes}

\firstpage{1} 
\pubvolume{1}
\issuenum{1}
\articlenumber{0}
\pubyear{2023}
\copyrightyear{2023}
\datereceived{ } 
\daterevised{ } %
\dateaccepted{ } 
\datepublished{ } 
\hreflink{https://doi.org/} %

\Title{Single-view 3D reconstruction via inverse procedural modeling}

\TitleCitation{Title}

\Author{Albert Garifullin $^{1,*}$\orcidA{},  Nikolay Maiorov $^{1}$\orcidB{} and Vladimir Frolov $^{1,2}$\orcidC{}}

\AuthorNames{Albert Garifullin, Nikolay Maiorov and Vladimir Frolov}

\AuthorCitation{Garifullin, A.; Maiorov, N.; Frolov, V.}

\address{%
$^{1}$ \quad Moscow State University, Moscow, Russia\\
$^{2}$ \quad Keldysh Institute of Applied Mathematics, Moscow, Russia}

\corres{Correspondence: albgar-14@yandex.ru}

\abstract{ We propose an approach to 3D reconstruction via inverse procedural modeling and investigate two variants of this approach. The first option consists in the fitting set of input parameters using a genetic algorithm. We demonstrate the results of our work on tree models, complex objects, with the reconstruction of which most existing methods cannot handle. The second option allows us to significantly improve the precision by using gradients within memetic algorithm, differentiable rendering and also differentiable procedural generators. In our work we see 2 main contributions. First, we propose a method to join differentiable rendering and inverse procedural modeling. This gives us an opportunity to reconstruct 3D model more accurately than existing approaches when a small number of input images are available (even for single image). Second, we join both differentiable and non-differentiable procedural generators in a single framework which allow us to apply inverse procedural modeling to fairly complex generators: when gradient is available, reconstructions is precise, when gradient is not available, reconstruction is approximate, but always high quality without visual artifacts. }

\keyword{single-view 3D reconstruction, procedural generation, inverse procedural generation, differentiable rendering}

\begin{document}

\section{Introduction}
Modern computer graphics application require an extensive range of 3D models and textures to populate virtual environments. These models can be created either manually, using the 3D modeling software, reconstructed from the real-life objects or created automatically with the procedural generators. 

Usually, some combination of these approaches is used, when generated or reconstructed models are fine-tuned by artists. It is especially common in video games, as every model in game is not only a 3D mesh with a set of textures, but also more complex object with levels of detail, animation, physics and so on. 3D reconstruction is a fundamental task in computer vision and hundreds of approaches were proposed in the past decades. High-quality 3D reconstruction of geometry can be achieved using the scanners or LIDARs and it's probably the best solution if all the equipment is available. Image-based approaches are much easier to use and will be the topic of this paper. In recent years, deep learning methods, such as NeRF \cite{nerf22}, have made remarkable progress solving the novel view synthesis task. While NeRF and it's successors are good at capturing details, they require significant amount of time and many input images. Other deep learning methods such as \cite{pontes2019image2mesh} and \cite{yang2022zeromesh} focus on single-view reconstruction, but their results are hardly applicable due to low quality. What is either important, is that deep learning methods rely on implicit representation of the model in the weights of neural network, and extracting the actual mesh and textures from becomes a challenging task \cite{nerfMeshing,nerfMeshing2}. 

Differentiable rendering is an another new and promising approach for 3D reconstruction, that is developing rapidly \cite{FujunDR21}. It facilitates gradient-based optimization of scene parameters, such as vertex positions and textures, to match the input images. Physical-based differentiable rendering can be used to reconstruct BRDFs for complex materials. Differentiable rendering can be used with implicit representations, such as signed distance functions (SDF) \cite{diffSDF}, but most approaches uses classic mesh-textures representation, which makes them easier to use than deep learning methods. 
Despite the progress made in recent years in image-based 3D reconstruction, there are still significant problems that prevent the wider use of these methods in real tasks. First is need many images of single object to perform good reconstruction (\cite{FujunDR21} uses 100 to achieve high quality). Single-view reconstruction methods are barely applicable, and others often require dozens of images from different views. Second is that even best approaches are struggling with reconstruction of complex objects with a lot of thin parts, such as trees and other vegetation. The key point here is that we don't actually need accurate reconstruction of all the small branches and leaves, we need a model to be detailed enough and look similar to the given reference. The third and perhaps the main problem of reconstruction remains that the models obtained as a result of its work are difficult to edit. Usually, the end result of such methods is one large triangular mesh. It does not store any information about the structure of the object, for example, which branch a particular triangle belongs to in the tree model. Also a big problem are various artifacts in the resulting model, to reduce the number of which various regularization or post-processing techniques are used.

While modern technologies for scanning and 3D reconstruction can produce high-quality models, they often require complex equipment and significant amounts of time. Additionally, it may be impossible to scan an object or obtain a sufficient number of images. For these cases, single-view 3D reconstruction methods are being developed. Deep learning methods \cite{pontes2019image2mesh,yang2022zeromesh,kerbl3Dgaussians} have made remarkable progress, but still have an issue: usually deep learning methods do not consider an important aspect of applications and human perception: structure and object components. Procedural generation is an alternative method for obtaining 3D models which consist of rules that establish a correlation between input parameters and 3D objects of a specific class. 

In this paper, we propose a novel approach that utilizes procedural generation for single-view and multi-view mesh and texture reconstruction. Instead of estimating the vertex positions of the mesh directly, we estimate the input parameters of a procedural generator through the silhouette loss function between reference and rendered images. By using differentiable rendering and creating partly differentiable procedural generators for gradient-based. Our approach also allows for convenient model modifications by changing the estimated input parameters, such as creating new levels of detail, altering the geometry and textures of individual mesh parts, and more.

\section{Related work}

\subsection{3D reconstruction}
Multi-view 3D geometry reconstruction has been well studied in the literature. Traditional and well-developed approaches for this task include structure from motion (SfM)\cite{schonberger2016structure} for large-scale and high-quality reconstruction and mapping (SLAM)\cite{cadena2016past} for navigation. These approaches are able to provide high quality results, but have two significant restrictions - first, they are unable to reconstruct unseen parts of the object, thus require a lot of different input images, and second, they fail working with non-lambertian (e.g. reflective or transparent) or textureless objects. These restrictions lead to the trend of resorting to learning based approaches, that usually consider single or few images, as it largely relies on the shape priors that it can learn from data. These methods use different scene representations, such as meshes \cite{wang2018pixel2mesh,nie2020total3dunderstanding,ye2021shelf}, voxel grids \cite{choy20163d,popov2020corenet}, point clouds \cite{fan2017point,chen2021unsupervised} or implicit functions \cite{chen2019learning}. Among these methods, mesh reconstruction is the most relevant to our work. The majority of single-view 3D mesh reconstruction methods employ an encoder-decoder architecture where the encoder extracts features from the image and the decoder deforms an initial mesh to the target shape. It is noteworthy that these methods are trained and evaluated on the same object categories. Further work \cite{tatarchenko2019single} showed that such approaches to single-view 3D reconstruction primarily perform recognition rather than reconstruction.

There are also a few research works focused on the generalized single-view 3D reconstruction \cite{zhang2018learning,yang2022zeromesh}, yet the quality of models reconstructed from a single image with these methods is often insufficient for practical use.

\subsection{Differentiable rendering and Neural SDF}
Physically-based differential rendering \cite{laine2020modular,zhang2021path,FujunDR21,FujunDR22,Bangaru2022NeuralSDFReparam} is an active area of research related to accurate 3D models and material reconstruction. Such algorithms provide a gradient of loss function with respect to the scene parameters that can be then minimized with the gradient descent. An appropriate scene representation is also important for this approach. Mesh-based representation has been the most widely studied, and specific regularization \cite{9577710} and modification for the Adam optimizer were proposed \cite{wang2021fast} for this task. Recently, Nicolet et al. combined recent advancements in this field in their work \cite{nicolet2021large} that significantly improved the quality of the resulting mesh. Other scene representations have also been studied. Vicini et al. \cite{vicini2022differentiable} proposed an algorithm for differentiable signed distance function (SDF) rendering and used it for multi-view reconstruction. Differential rendering can also be combined with deep learning \cite{yang2022zeromesh} to provide face quality supervision and regularization. The main limiting factor for these approaches is the requirement for a large number of input images (in order of 100) which is difficult in practice.

Special mention should be made of the work \cite{convexphd} which, to our option, is a kind of precursor for differentiable rendering based approaches. This work compute a surface with minimal area that best fits the input data and suitable priors as the solution of a variational, convex optimization problem.

\subsection{NeRF}
Mesh-based or surface-based representations enable efficient rendering and produce easy-to-use meshes as a result of their work. However, image-based optimization of surface geometry can be challenging due to the non-convexity of such an optimization. Volumetric representations \cite{lombardi2019neural,vicini2021non} can reliably reach a desirable minimum, but are usually unable to capture finer details. Alternatively, point-based shape representations have also been shown to produce high-quality scene reconstructions \cite{yifan2019differentiable,ruckert2022adop}.
Another key tool for scalable scene representation is the use of coordinate-based neural networks, also known as neural fields \cite{mildenhall2021nerf,fridovich2022plenoxels,muller2022instant}, which push beyond the resolution limits of discrete grids and generalize to higher-dimensional signals, such as directional emission \cite{mildenhall2021nerf}. Neural fields \cite{muller2022instant} have demonstrated the ability to handle complex scenes and produce compelling results within seconds. However, to utilize the obtained results, one typically needs to convert them from an alternate representation into a 3D mesh, as most rendering engines work primarily with meshes. Such transformations can pose a significant challenge, resulting in suboptimal models with a large triangle count and lower visual quality \cite{muller2022instant}. Another serious NeRF drawback is the same as for differential rendering based approaches -- a large number of input images is required. Compared to pure differentiable rendering approaches, NeRF-based approaches are more stable and don't require specific initial conditions for optimizations, but yield models which much more hungry by memory consumption and computational power that is needed for their direct rendering.

\subsection{Procedural modeling}
Procedural content generation is a well-known approach to creating diverse virtual worlds. It is used on different levels, ranging from individual objects to large-scale open worlds and game scenarios \cite{hendrikx2013procedural,freiknecht2017survey}. Many procedural object generators work as functions that transform a vector of numerical parameters into a 3D mesh of a specific class, e.g., trees, buildings. 
Procedural modeling of trees poses a particular interest, as it is especially hard and tedious to create detailed vegetation manually. Projects like SpeedTree \cite{SpeedTree} provide an artist with a tool that allows him to interactively create highly detailed tree models, while autonomous procedural generators \cite{prusinkiewicz2012algorithmic,yi2018tree} create a model of a plant without human involvement, based on a set of input parameters. Due to its complexity, it is much harder to perform a traditional 3D reconstruction of a tree model, as many details are hidden by the foliage. Considering this, as well as the fact that procedural generators are able to create high-quality tree models, a few specialized methods for tree reconstruction using procedural generators have been proposed. 

\subsection{Inverse procedural modeling}

Work \cite{stava2014inverse,urban1} proposes an inverse procedural generation. i.e. selecting parameters of generator based on input 3D model (i.e. takes polygonal tree models as input) of a tree to recreate it with the generator. Sampling of trees were implemented via Markov Chain Monte Carlo in parameter space. The main limitation of this work is that it already requires a 3D mesh at the input. SIGGRAPH course \cite{Aliaga16SiggCourse} provides a comprehensive review of inverse procedural modeling based on already existing 3D content. Procedural-based reconstruction based on a single or few images, however, remains an open problem in general.

In order to move from 3D modeling to a 3D reconstruction problem, it is necessary to propose a similarity metric which answers a question: ``To what extent does a given set of procedural generator parameters allow one to obtain a tree model similar to the input reference?'' 

This way was adapted in \cite{guo2020inverse} and \cite{garifullin2022fitting}. Work \cite{guo2020inverse} analyzes the input image of tree and infers a parametric L-system that represents it. \cite{garifullin2022fitting} uses more complex stochastic procedural models which simulate vegetation growth process and therefore carried out the selection of procedural generator parameters using genetic algorithms. Both methods takes a single image of a tree and find an optimal generator's parameters by minimizing the loss function between the given reference and image of the model created with the generator. Despite relatively good results, these methods are specific for plants (especially \cite{guo2020inverse}) and cannot be applied to a more broad class of objects. 

There are some works that focus on inverse procedural modeling of specific classes of objects, such as materials \cite{hu2022inverse}, facades \cite{wu2013inverse} (which  derives a so-called ``split grammar'' for a given facade layout) and knitting yarns \cite{FujunFiber2016,trunz2023neural}. More general approach has also been studied. In the work \cite{gaillard2022automatic} the authors focus on interactive tool for procedural generation to increase the user's control over the generated output. Then the user modifies the output interactively, and the modifications are transferred back to the procedural model as its parameters by solving an inverse procedural modeling problem.  

\subsection{Adding geometry constraints}
There are a number of methods focusing on the reconstruction of bodies of revolution \cite{sv3d0}, \cite{sv3d1} , buildings \cite{sv3d_buildings}, human bodies \cite{sv3d_humans,sv3d_humans2} and e.t.c. These approaches are carefully designed for specific application scenarios and input data. They may \cite{sv3d_buildings} or may not use learning-based approaches. Although such approaches usually provide sufficiently accurate results for their intended applications. However, they are more limited and specific to a certain type of object than the procedural approach in general.

\subsection{Motivation for our method}

Our work is located at the intersection of 3D reconstruction and inverse procedural modeling. On the one hand, we wanted to create a reconstruction method that would be devoid of visual artifacts, which is quite common problem for existing solutions when low amount of input images are available. On the other hand, we wanted to create an image-driven procedural models which allow artists to create not just a single, but a family of models with desired properties: existing methods takes 3D model as input \cite{stava2014inverse,urban1} or either require additional manual user input \cite{gaillard2022automatic}. Our approach is more general because it takes as input images that can be obtained in a variety of ways, including generative neural networks. That is why the requirement of single input image is so important for us.

\section{Method Overview}
In this section we describe our method for single- and multi-view 3D reconstruction using the procedural generation. First we present our general reconstruction pipeline and two variants of it - differentiable and non-differentiable ones. The advantages and limitations of differentiable variant are presented. Then we present procedural generators used in our work and discuss the differences between these generators in case of using them for reconstruction. Next part of this section is dedicated to the optimization process both in differentiable and non-differentiable pipelines. 

\subsection{Pipeline}
In our work we assume than we are given a set of $N$ images $R_{i}$ of the same object from different (and unknown) viewpoints. As our goal is geometry reconstruction, we do not use the reference images directly and instead use "masks" $M_{i}$ taken from these images. The exact look of the mask depends on the procedural generator, but in most cases it is as simple as binary mask that separates object from background. We assume that we already know the procedural generator $Gen$ needed for reconstruction, as it was chosen manually or with some sort of classification. It takes a set of parameters $P$ to create a triangle mesh consisted of vertex positions, normals and texture coordinates $Gen(P) = Mesh = \{pos_i, norm_i, tc_i\}$ where $pos_i$ -- vertex position, $norm_i$ -- vertex normal and $tc_i$ -vertex texture coordinates of $i$-th vertex. Then the mask of this mesh can be created with the render function that takes only vertex positions $pos$ and camera parameters $C$: $I = Render(Mesh, C)$. We are performing the reconstruction as an optimization of loss function between given and rendered masks:
\begin{linenomath}
	\begin{equation}
	Loss(P, C_1, ..., C_N) = \frac{1}{N} * \sum_{i=1}^{N}MSE(I_i, M_i) = \frac{1}{N} * \sum_{i=1}^{N}MSE(Render(Gen(P), C_i), M_i)
	\end{equation}
\end{linenomath}

\begin{figure}[H]
	\includegraphics[width=13 cm]{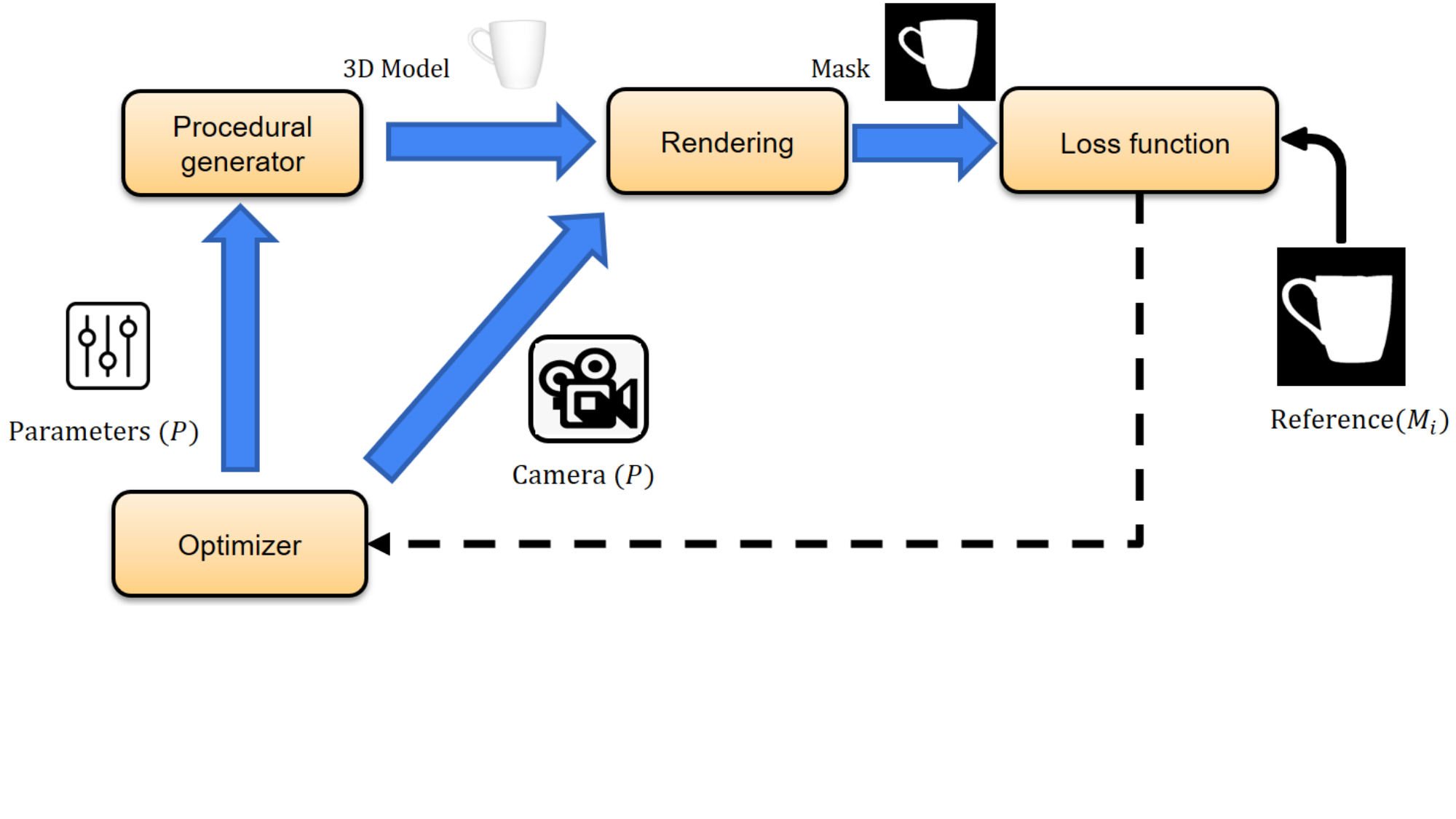}
	\caption{The mesh reconstruction cycle consists of a procedural generator creating a mesh from input parameters, which is then rendered as a mask and loss function is then calculated between it and the mask of reference image. The loss value is then returned to an optimizer that updates generator's and cameras' parameters.
		\label{fig3}}
\end{figure}

Then the goal of our algorithm is to minimize the loss function and find optimal parameters of generator $P^*$ and cameras $C^*_1, ..., C^*_N$. It brings us to the general reconstruction pipeline shown on the Figure~\ref{fig3}.

This general reconstruction pipeline looks plain and simple, especially because it treats the procedural generator as a black-box function that is able to create a 3D model from a set of parameters. This fact allows us to use any existing procedural generator without knowing how it actually works inside, at least in theory, as we still need to know which input parameter values are correct and which are not. However, in this case we have to implement gradient-free optimization method for loss function (1), that can pose a significant challenge. Later we will describe the cases where we have to use this general pipeline and show how to make optimization for it relatively effective, but for most cases we can make a few changes to this pipeline that would allow gradient-based optimization. Doing so requires gradients of the loss function with respect to scene parameters. Using differentiable renderer for mask rendering and loss function calculation, we can obtain $\frac{dLoss}{dC}$ where $C$ is camera parameters. Our implementation uses Mitsuba3  for this purpose \cite{Mitsuba3}. Obtaining 
$\frac{dLoss}{dP}$ (where $P$ is procedural model parameters) is a bit more complicated. Considering the chain rule and assume that 
the mask does not depend on model's normal vectors and texture coordinates, we get:
\begin{equation}
	\frac{dLoss}{dP} = \frac{dLoss}{dpos} * \frac{dpos}{dP}
\end{equation}

\begin{figure}[H]
	\includegraphics[width=13 cm]{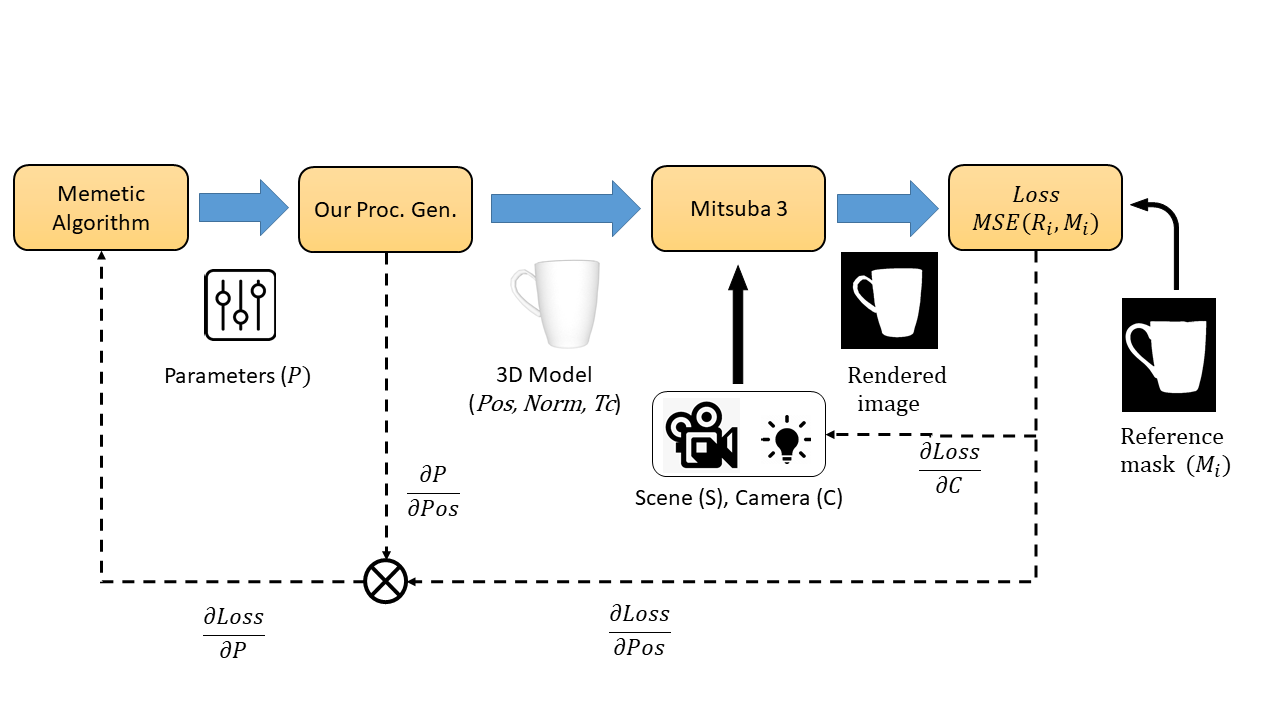}
	\caption{TODO: FIX IMAGE TO MATCH THE GENERAL PIPELINE
			 Gradient-based reconstruction pipeline. Rendering and loss function calculation are both done by the differentiable renderer. Assuming that the rendered mask depends only on vertex positions $pos$, but not normals and texture coordinates, we get $\frac{dLoss}{dC}$ and $\frac{dLoss}{dpos}$. Then to get $\frac{dLoss}{dP}$, the procedural generator should be able to calculate jacobian $\frac{dpos}{dP}$ alongside with the mesh $(pos, norm, tc)$. 
		\label{fig4}}
\end{figure}

Equation term $\frac{dLoss}{dpos}$ (where $pos$ is a vector of all vertex positions) also comes from differentiable renderer and jacobian $\frac{dpos}{dP}$ needs to be obtained from procedural generator. The whole process of optimizing the silhouette loss function is illustrated in Figure~\ref{fig4}. Making the procedural generator that is able to calculate $\frac{dpos}{dP}$ is the most important part of this improved pipeline. We would call a generator that is able to do it differentiable procedural generator. In the next section we will describe differentiable procedural generators created for this work and challenges that they pose. In fact, most non-trivial generators can be only partly differentiable and we address this issues in the third section, where the optimizer is described in details.  

\subsection{Differentiable procedural generators}
The pipeline described in the previous section treats the procedural generator as a black-box function that can produce a mesh $(pos, norm, tc)$ (where $pos$,$norm$ and $tc$ are vectors of all positions, normals and texture coordinates) and jacobian $\frac{dPos}{dP}$ from an input vector $P$. In practice, it is impossible to make a generator of non-trivial objects that works as a smooth, differentiable function.  This is primarily due to objects having discrete properties, such as the number of floors in a building. For consistency, we still include these parameters in the vector $P$, but for each parameter $P_d$ of this type, we assume that $\frac{dPos}{dP_d} = 0$. \par
\begin{figure}[H]
	\includegraphics[width=13 cm]{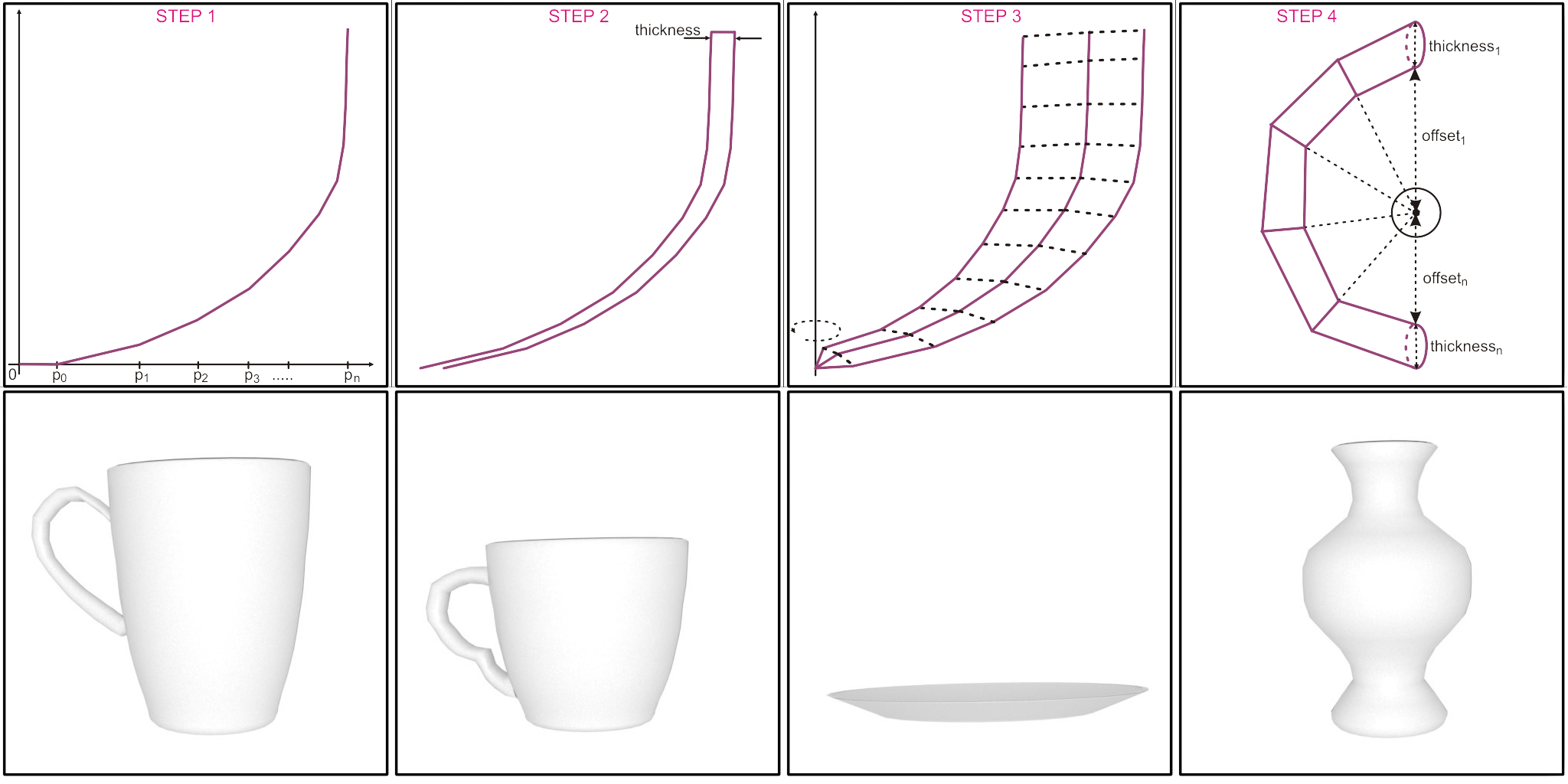}
	\caption{
		Dishes generator. Top row: generation steps. 1) Create a spline from a vector of vertical offsets. 2) Transform it to create a closed spline with thickness. 3) Rotate it to get the dish body. 4) (Optional) Create a handle from a circle spline and a vector of offsets. The number of points in splines can be changed to produce different levels of detail. Bottom row: some examples of generated dishes. 1) Mug 2) Tea cup 3) Bowl 4) Jar
		\label{fig5}
	}
\end{figure}
\begin{figure}[H]
	\includegraphics[width=13 cm]{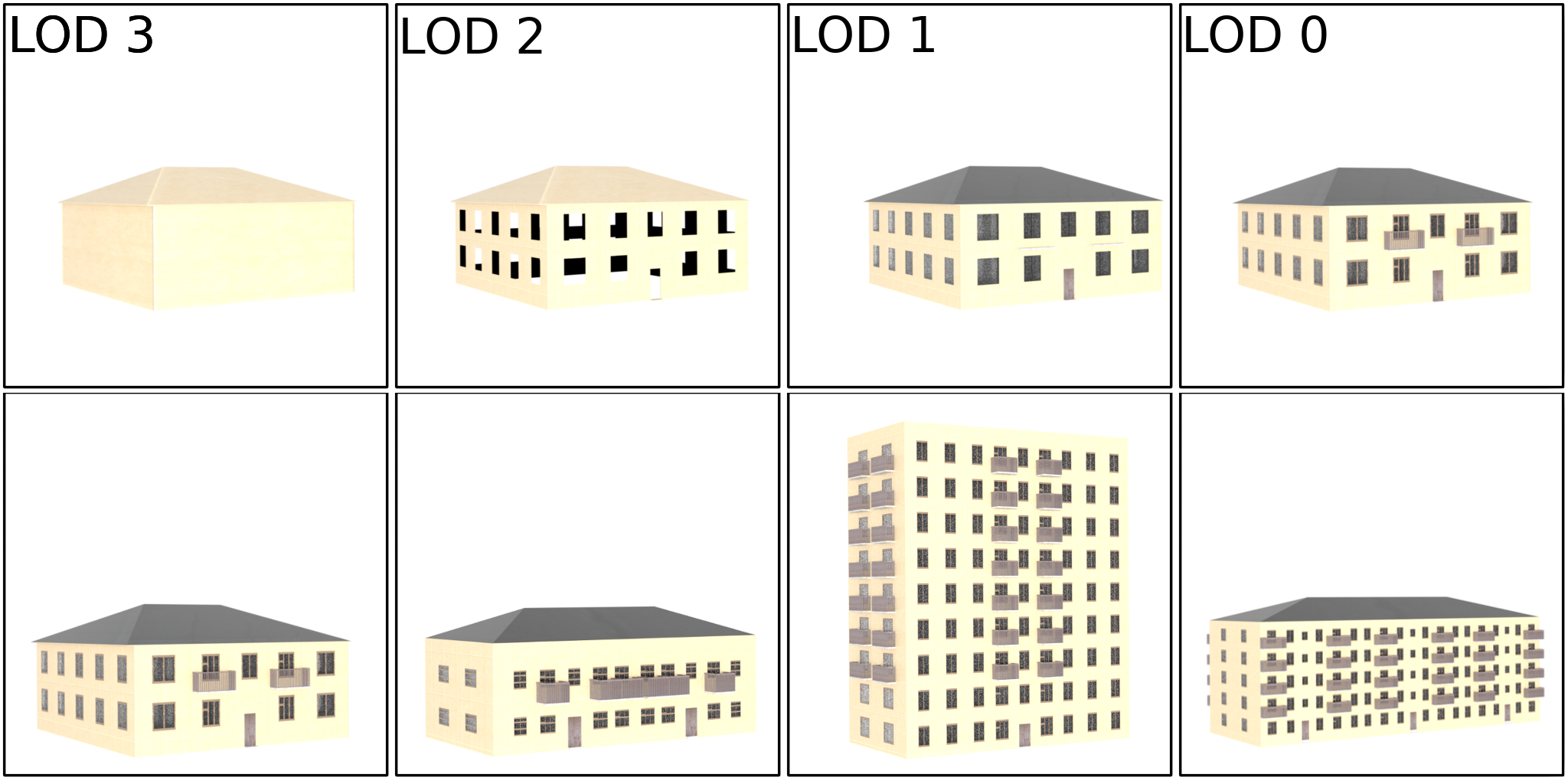}
	\caption{
		Building generator. Top row: Levels of detail for the same building. 1) Box with a roof. 2) Only outer walls. 3) Building without small details 4) Full detailed building
		Bottom row: some examples of generated buildings
		\label{fig6}
	}
\end{figure}

For this work, two different procedural generators are created: for dishes and buildings. Both were developed from scratch using the CppAD automatic differentiation library \cite{bell2012cppad} to obtain the required derivatives. The dishes generator (Figure~\ref{fig5}) is an example of a simple and easy-to-differentiate algorithm that only has one binary parameter. In contrast, buildings have numerous discrete features, and the building generator (Figure~\ref{fig6}) was created to demonstrate the ability of our method to handle such challenges.

\subsection{Gradient-based Optimization}
Even simple procedural models from previous section prove to be extremely challenging functions to optimize due to two factors: the large number of non-linear internal dependencies in the generator and heterogeneous input parameters, some of which are integers with a limited set of possible values. Previous efforts at 3D reconstruction with procedural generators \cite{garifullin2022fitting} relied on a specific implementation of the genetic algorithm \cite{mitchell1998introduction} to find solutions without gradient calculation, but only for a restricted set of problems. Being able to calculate the gradient of the loss function expands the list of available methods for optimization. However, we still cannot obtain the derivatives with respect to parameters that represent the procedural model's discrete features.

\begin{figure}[H]
	\includegraphics[width=13 cm]{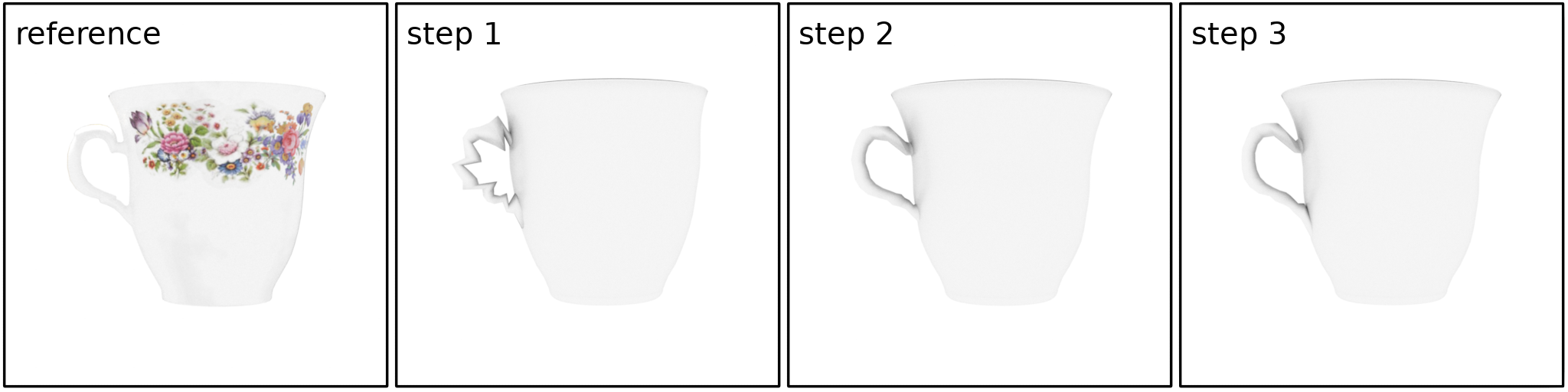}
	\caption{\label{fig:optimization}
		Results of model reconstruction after each optimization step. From left to right: reference image, genetic algorithm with 128x128 rendering resolution, gradient descent with resolution of 256x256 and 512x512 respectively}
\end{figure}

To address this issue, we use the memetic algorithm \cite{neri2012memetic}, which combines genetic algorithm with gradient-based optimization. We start with an initial population, a set of initial parameter values, which is taken from presets that have been prepared in advance. Each preset is a set of input parameters representing an adequate model (like those shown on Figure~\ref{fig5}). The memetic algorithm performs random mutations and recombination of the initial population, along with gradient-based optimization. Although this process requires several thousand iterations, it can be performed on models with low level of detail and low rendering resolution. To further improve the quality, we use the solution obtained from the memetic algorithm as an initial approximation for the next round of gradient-based optimization with higher levels of detail and higher rendering resolution. For texture reconstruction, only the gradient-based optimization step is performed. The results at each step are shown in Figure~\ref{fig:optimization}.

\subsection{Non-differentiable Case}
Using differentiable rendering and procedural generation allows us to use gradient descent and makes optimization a lot easier even with integer-valued input parameters that should be optimized differently. However, some types of procedural  generators do not fit to the described pipeline. Modern systems for automatic modeling of plants, such as \cite{hadrich2017interactive}\cite{yi2015light}\cite{yi2018tree} concentrate on simulating the growth process and the influence of the environment on it. Such simulation is crucial for more realistic output, but  at the same time it makes the creation process more random and less stable. We implemented a generator based on work \cite{yi2018tree} and found out that even a tiny change in input parameters for it can lead to a significantly different tree structure. The overall tree shape stays the same, but the exact location of specific branches changes. In this case it is extremely hard to obtain meaningful gradients with automatic differentiation. The same problem to some extend also appeared in the much simpler rule-based generator \cite{weber1995creation}. So the trees modeling shows the limitations of differentiable reconstruction approach but it is still possible to perform the reconstruction using the general pipeline (Figure~\ref{fig3}). As we cannot obtain the gradients of loss function in this case, we used a special version of the genetic algorithm. Also a special loss function was proposed for reconstruction of tree models.

\subsubsection{Tree Similarity Function}
\begin{figure}[H]
	\includegraphics[width=13 cm]{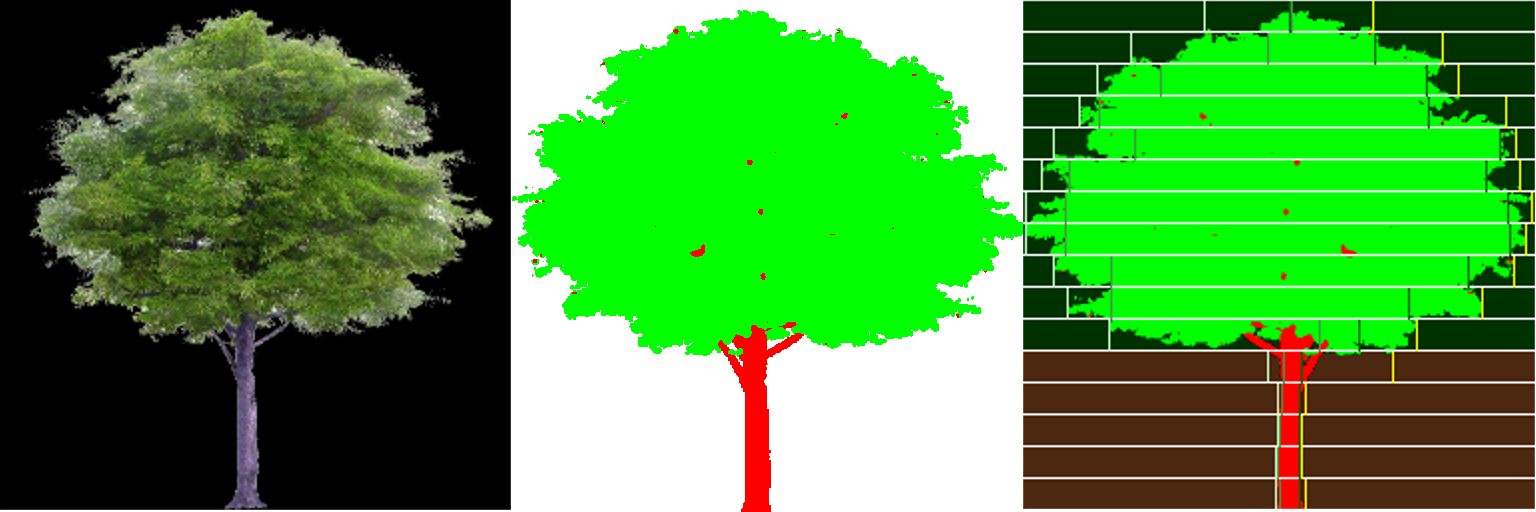}
	\caption{\label{fig_tree_1}
		The original image, semantic mask, visualization of the division into stripes. Stripes of brown color correspond to the trunk, green color to the crown. Vertical lines inside each stripe - values $a_i$, $b_i$, $c_i$, $d_i$ respectively}
\end{figure}

\begin{figure}[H]
	\includegraphics[width=13 cm]{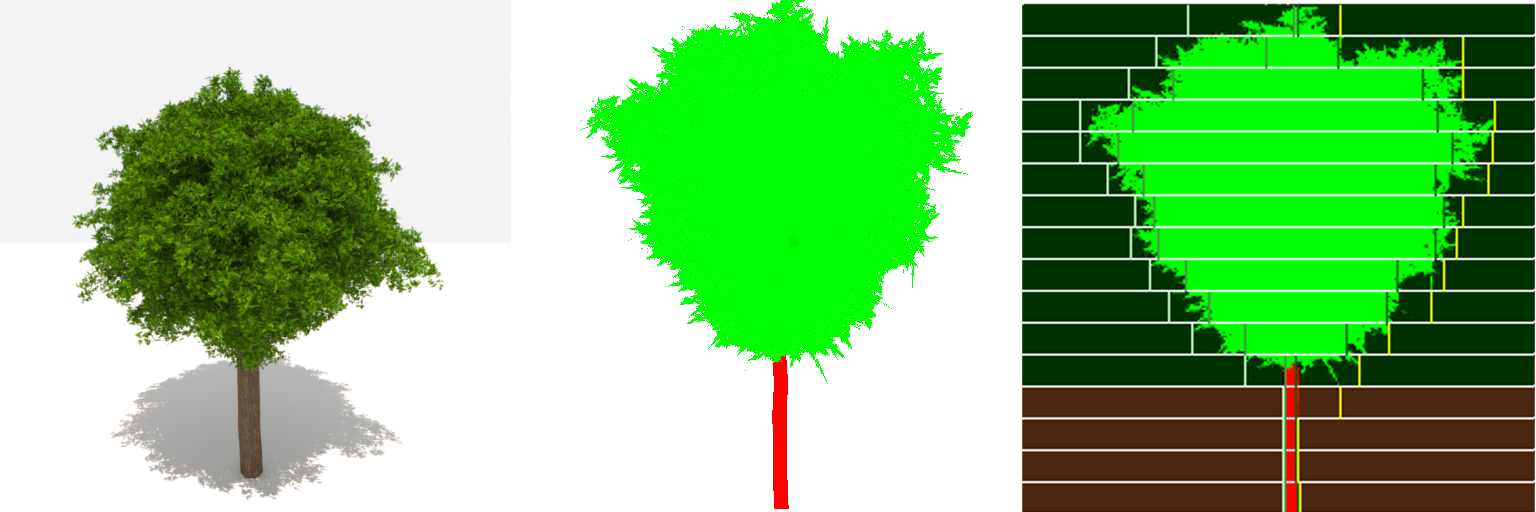}
	\caption{\label{fig_tree_2}
		Generated tree, semantic mask (imposter), visualization of the division into stripes}
\end{figure}

We define the loss function between generated tree model an reference image through the tree similarity function and an optional set of regularization multipliers.  
\begin{linenomath}
	\begin{equation}
		Loss = TSim * \prod_{c \in C} \frac{min(c, c_{ref})}{max(c, c_{ref})}
	\end{equation}
\end{linenomath}
Tree similarity value is obtained by comparing the reference image with the impostors of the generated tree model. Semantic masks are used for comparison, where each pixel belongs to one of three categories: branch, foliage, background. To obtain such a mask from the source image, a neural network can be used similarly to \cite{li2021learning}, and in simple cases, we can get it based only on pixel color (green corresponds to leaves, brown or gray - to branches and trunk). The image is divided into 20-30 narrow horizontal stripes, for each of which are determined:
\begin{itemize}
	\item	$[a_i, d_i]$ - crown borders
	\item	$[b_i, c_i]$ - dense crown borders (>75\% leaves pixels)
	\item	$B_i$ - branches pixels percentage
	\item	$L_i$ - leaves pixels percentage
\end{itemize}
According to the ratio $B_i/L_i$, each stripe refers either to the crown or to the trunk, see Figure~\ref{fig_tree_1} and Figure~\ref{fig_tree_2}. Comparing the parameters $a_i$,$b_i$,$c_i$,$d_i$ and $B_i/L_i$ ratio for every stripe of the reference image and the image of the generated model, we calculate the value of tree similarity $TSim$. 


Regularization multipliers $\frac{min(c, c_{ref})}{max(c, c_{ref})}$ are used to guide the optimization to the result, not only close to reference by $TSim$ but also with desired characteristics. Among them may be:
\begin{itemize}
\item	Number of vertices in the branch graph
\item	Height and width in some scale
\item	Average branches and leaves density
\item	Average leaf size
\end{itemize}	
None of these characteristics is mandatory, but it is recommended to specify the number of vertices in the branch graph, otherwise the search for a solution will slow down due to the need to search for it among models with very high geometric complexity.

\subsubsection{Genetic Algorithm Implementation}
\begin{figure}[H]
	\includegraphics[width=9 cm]{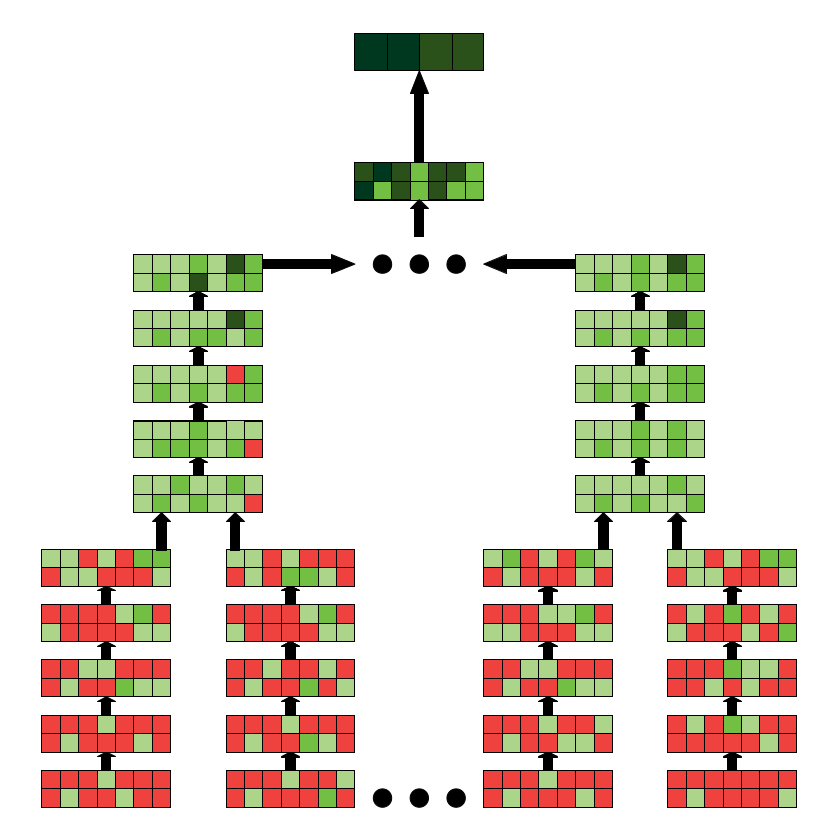}
	\caption{\label{fig_tree_3}
		Tree-like population structure. Zero-level elementary GA are started with random genes, while GA from level i takes best species from two final populations of the previous level. The final result of a whole algorithm is a small set of best species on the very top level}
\end{figure}

The previously mentioned similarity function is used as an objective function for the genetic algorithm. 
A proposed genetic algorithm consists of several elementary genetic algorithms, with a selection of the best results of each of them. Each elementary GA includes the initialization of a population and its evolution over a fixed number of generations. In the figure, each vertex of the tree is such an elementary GA. Algorithms on the leaves of the tree start with a randomly initialized population, and all the others form a population of the fittest “individuals” obtained in the child vertices. Figure~\ref{fig_tree_3} shows proposed tree-like population structure.\par

All elementary GA work according to the same strategy ($f(x)$ - objective function). At the beginning of each iteration, half of the population with the worst fitness value is removed. The remaining individuals take part in the creation of a new generation. At each of the vacant places in the population, a new individual is created with one-dot crossover, its parents are selected from the remaining species of the previous generation. The fitness proportionate selection is used, which means that the probability of choosing an individual as a parent is proportional to the value of its fitness. For representatives of the new generation, the values of the objective function and the fitness function are calculated.\par

Mutation chance $M_{chance}$ and percentage of genes to change  $M_{genes}$ are constant. 
Here is a proposed method for genome $G$ mutation:
\begin{enumerate}
\item	Values of $n*M_{genes}$ randomly chosen genes are changed. The probability of mutation in the $k$ gene is proportional to the average rate of change of this gene $V(k)$
\item	For the result genome $G'$ we estimate its quality $Q(G')$
\item	Steps 1-2 are repeated several times (500 in the experiment) and the mutation results in a gene with a better quality score
\end{enumerate}
Functions  $V(k)$ and $Q(G')$ based on pre-collected information about the entire family of objective functions $F = {f(x)}$ (each function corresponds to its own input image and set of properties) on a large sample with random input parameters $x$.
\begin{linenomath}
	\begin{equation}
		V(k) = \mathrm{\frac{f(x+h*x_k) - f(x)}{h}}
	\end{equation}
\end{linenomath}
\begin{linenomath}
	\begin{equation}
		P_{i,j} = P(F(x) > \epsilon | \frac{j-1}{B} < x_i < \frac{j}{B}) 
	\end{equation}
\end{linenomath}
\begin{linenomath}
	\begin{equation}
		P_0 = P(F(x) > \epsilon)
	\end{equation}
\end{linenomath}
\begin{linenomath}
	\begin{equation}
		Q_{i,j} = sign(P_{i,j} - P_0) * \frac{max(P_{i,j}, P_0)}{min(P_{i,j}, P_0)}
	\end{equation}
\end{linenomath}
\begin{linenomath}
	\begin{equation}
		Q(G) = \sum_{i=1}^{N}Q(i, \left \lfloor G_i * B \right \rfloor)
	\end{equation}
\end{linenomath}
$N$ is a number of input parameters, while $\epsilon$, $n$, $h$ and $B$ are hyper-parameters of our implementation of GA.

\section{Results}
In the following, we demonstrate the optimization results using our method, compare it with other approaches and discuss some details of implementation. 

\subsection{Implementation}
We implemented our method as a standalone program written mostly in C++. Differentiable procedural generators were written from scratch and CppAD library was used for automatic differentiation inside generators. Mitsuba 3 was used for differentiable rendering and also to obtain images shown in this section. We also implemented differentiable renderer for fast rendering of silhouettes during the optimization. For some models in this section we performed texture reconstruction using Mitsuba 3. \par
Our method requires defining a set of hyperparameters, such as number of rendering stages, render image size, number of iterations for each stage and parameters of genetic or memetic algorithm. We used mostly the same set of hyperparameters for all our results. For reconstruction with differentiable generators we set number of optimization stages is 3 or 4, with rendering resolution 128x128 for the first stage and it doubles on each next stage. Memetic algorithm is used only on a first stage with limit of 5000 iterations, for next stages Adam optimizer is used with 200-300 iterations per stage. For reconstruction with non-differentiable trees generators we used only one stage with 256x256 rendering resolution and 50 thousand iterations limit for genetic algorithm. The whole optimization process takes about 10-20 minutes with differentiable procedural generator and up to 30 minutes with non-differentiable one. These timings have been measured on PC with AMD Ryzen 7 3700X and NVIDIA RTX 3070 GPU and notebook with Intel Core i7 7700HQ and NVIDIA GeForce GTX 1060. 

\subsection{Optimization Results}
\begin{figure}[H]
	\includegraphics[width=13 cm]{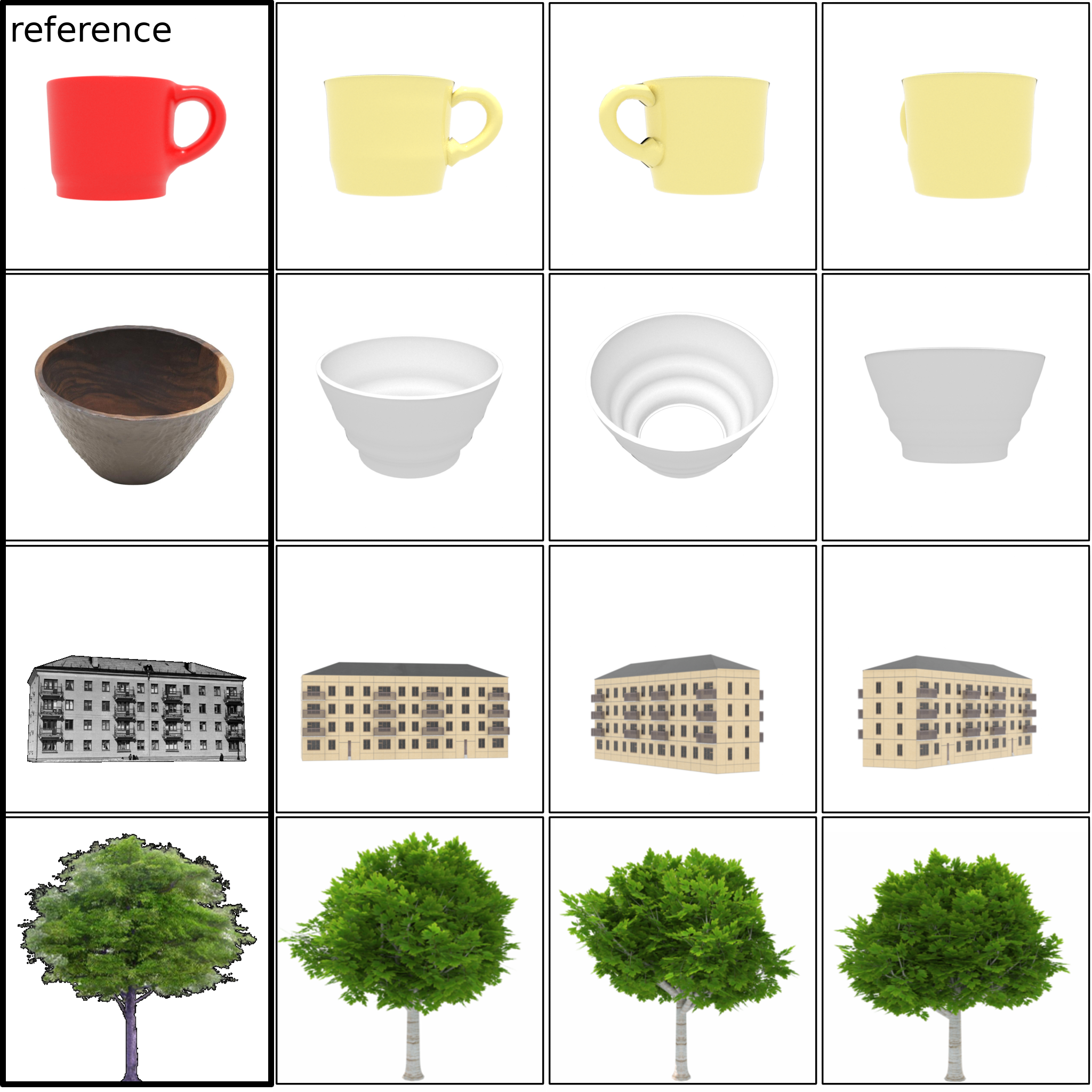}
	\caption{\label{fig_res_1}
		Reconstruction results for different types of objects. Differentiable procedural generators were used for cups and buildings and non-differentiable for trees. All models were rendered with some default textures with the texture coordinates provided by procedural generator.}
\end{figure}
Figure~\ref{fig_res_1} shows results of single-view reconstruction for different types of objects using the corresponding procedural generators. For reconstruction of buildings we provided manually created masks for windows, while other masks were made automatically.  As expected, all presented models are free of usual reconstruction artifacts, are detailed and have adequate texture coordinates that allow to apply textures on them. Reconstruction of cups and building with differentiable generators results in models very close to reference. The main quality limitation here is the generator itself. The quality of reconstruction with non-differentiable tree generator is lower, as genetic algorithm struggles to find local minimum precisely. However, the overall shape of the tree is preserved and the model itself is good enough.

\subsection{Comparison to Other Approaches}
\begin{figure}[H]
	\includegraphics[width=11 cm]{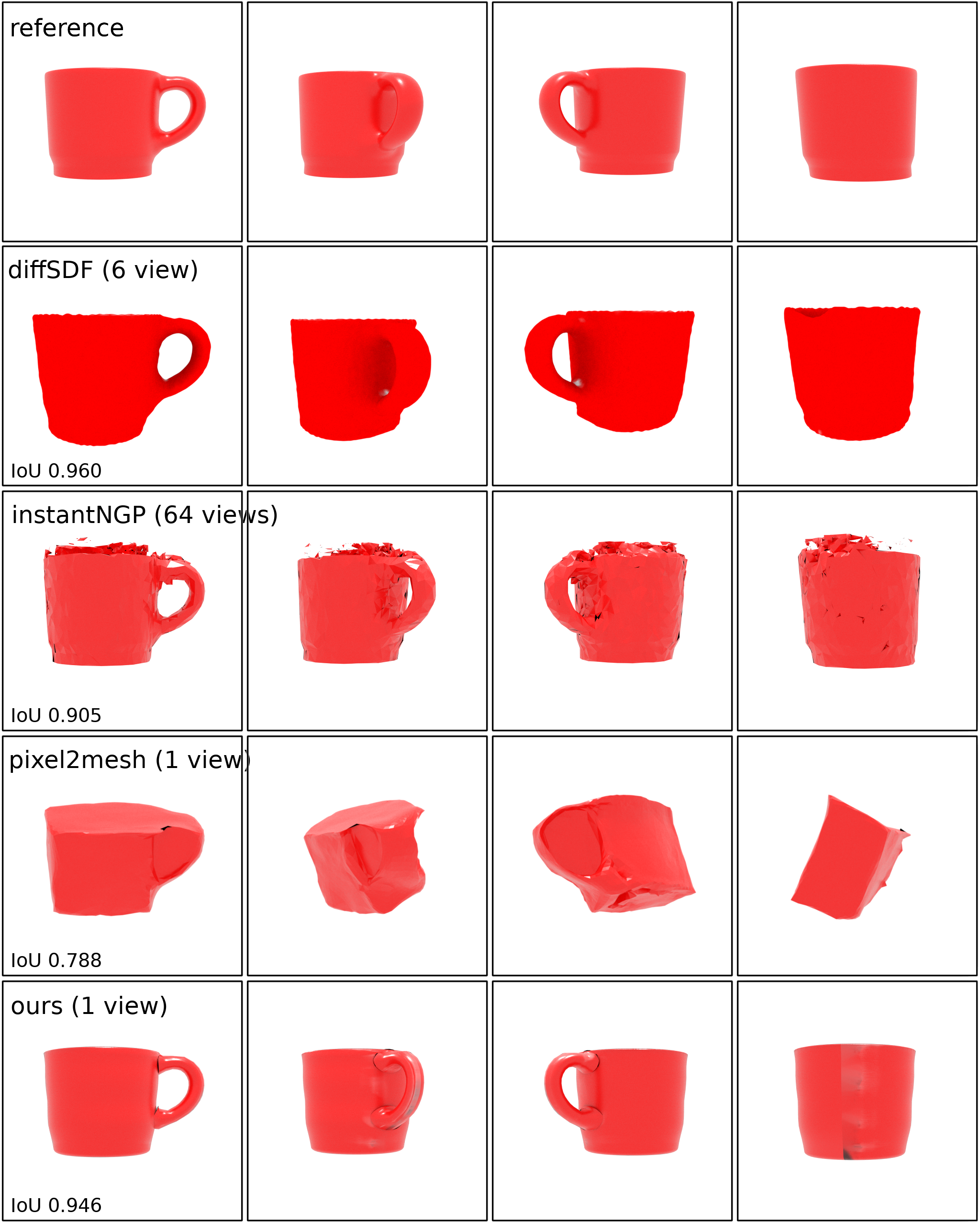}
	\caption{\label{fig_res_2}
		Results using our method compared to differentiable SDF reconstruction, InstantNGP and Pixel2Mesh. We measured the average silhouette intersection over union (IoU) between reference and reconstructed models for 64 uniformly distributed viewpoints. Our approach is far better than Pixel2Mesh single-view reconstruction and has comparable results to multi-view reconstruction methods.}
\end{figure}
\begin{figure}[H]
	\includegraphics[width=13 cm]{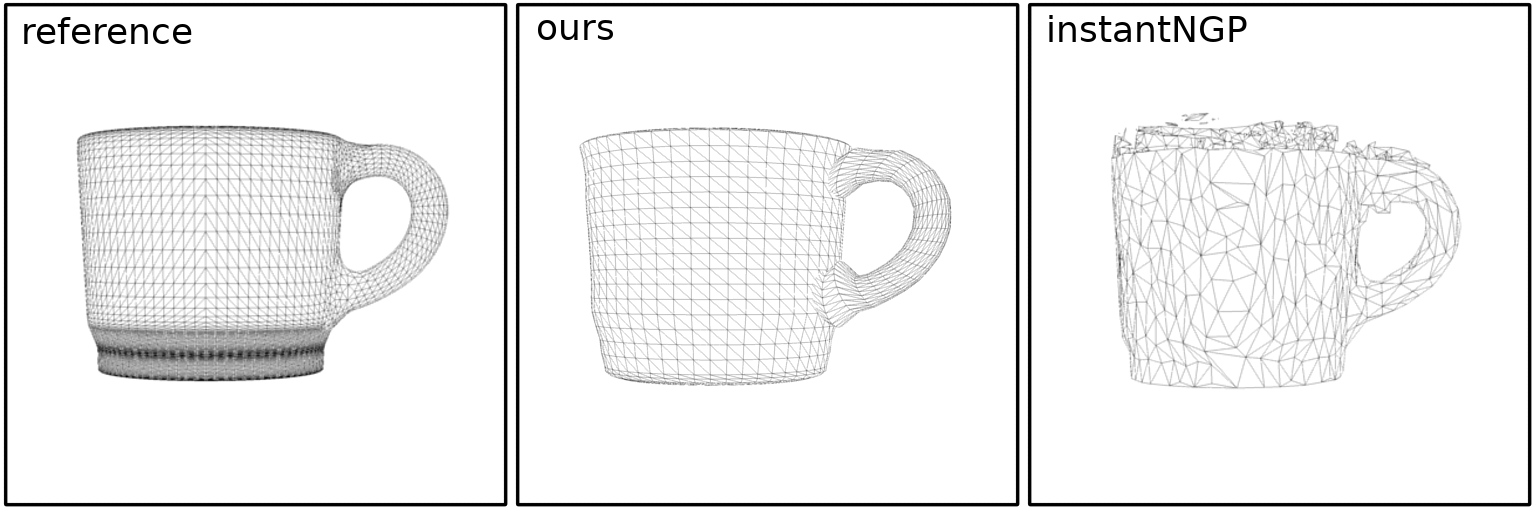}
	\caption{\label{fig_res_3}
           Wireframe meshes comparison. Our method is able to represent model with less triangles and smaller number of tiny and malformed polygons, which allows to use it directly in applications.}
\end{figure}
\begin{figure}[H]
	\includegraphics[width=13 cm]{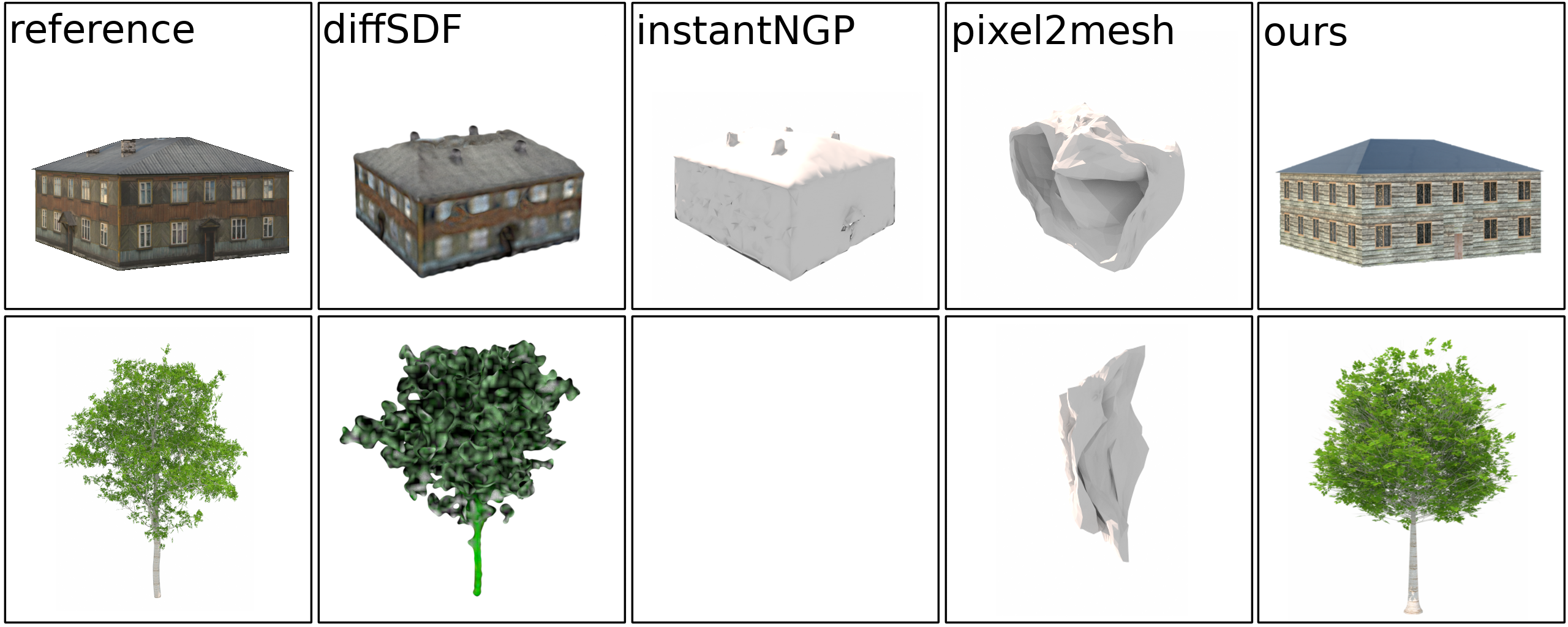}
	\caption{\label{fig_res_4}
		Building and tree reconstructed with our approach, differentiable SDF reconstruction, InstantNGP and Pixel2Mesh. InstantNGP failed to reconstruct the tree model and no mash can be obtained from the radiance field it provided.}
\end{figure}

We compared results of our work with different modern approaches to image-based 3D reconstruction to demonstrate how our method allows to avoid issues common for all this algorithms. Three algorithms were used for comparison: DiffSDF \cite{vicini2022differentiable}, a method that represents the scene an a signed distance field and utilizes differentiable rendering for it's optimization; InstantNGP \cite{muller2022instant} is an state-of-the-art NeRF-based approach from NVidia  and Pixel2Mesh \cite{wang2018pixel2mesh} is a deep neural network able to produce 3D mesh from a single image. Figure~\ref{fig_res_2} demonstrates the results of different approaches on a relatively simple model of a cup. It shows that that from Pixel2Mesh failed to reconstruct the shape of the model from one image, while other methods used multiple images and achieved the result comparable to ours. However, they both struggled to represent concave shape of the model and created meshes with substantial artifacts. On the Figure~\ref{fig_res_3} reconstructed meshes are rendered in wireframe mode.
For more complex objects, such as trees and buildings, the ability to reconstruct object's structure becomes even more important. For most applications, such as video games, reconstructed meshes have to be modified and augmented with some specific data and it becomes much easier if the initial mesh has some structural information. In this particular case it is necessary to distinguish between triangles related to different parts of the building or to the leaves/branches of a tree in order to be able to meaningfully modify these models. Figure~\ref{fig_res_4} shows results of buildings and trees reconstruction with different approaches. 
We also tested our approach and other on more models from given classes and compared it with diffSDF, InstantNGP and Pixel2Mesh using different number of input images. The full results of the comparison on all the studied models are shown in the Table~\ref{res_table} below. Despite the fact that our method, on average, shows worse results than diffSDF and InstantNGP by IoU metric, it is able to create better models even when it is not possible to achieve a high similarity to the reference images. This is especially noticeable in Figure~\ref{fig_res_4}, where the diffSDF method create the model closer in shape to the original, but failed to reconstruct the expected structure of the tree.

\begin{table}[H] 
	\caption{Average IoU on different models.Larger is better. "-" is put in cases, where the method failed to reconstruct the model at least remotely similar to the reference mesh. \label{res_table}}
	\newcolumntype{C}{>{\centering\arraybackslash}X}
	\begin{tabularx}{\textwidth}{CCCCCC}
		\toprule
		\textbf{Method}	& \textbf{cup\_1}	& \textbf{cup\_2} & \textbf{building} & \textbf{tree\_1} & \textbf{tree\_2}\\
		\midrule
		diffSDF(2 views)		& 0.787 & 0.422 & 0.728 & 0.665 & 0.746\\
		diffSDF(6 views)		& 0.960 & 0.502 & 0.967 & 0.876 & 0.911\\
		diffSDF(12 views)		& 0.984 & 0.669 & 0.979 & 0.900 & 0.941\\
		instantNGP(16 views)	& 0.858 & 0.878 & 0.940 &   -   &   -  \\
		instantNGP(64 views)	& 0.905 & 0.930 & 0.971 &   -   &   -  \\
		pixel2Mesh(1 view)		& 0.788 & 0.513 & 0.626 &   -   &   -  \\
		ours                	& 0.940 & 0.967 & 0.886 & 0.509 & 0.573\\
		\bottomrule
	\end{tabularx}
\end{table}

\subsection{Multi-view Reconstruction}
\begin{figure}[H]
	\includegraphics[width=13 cm]{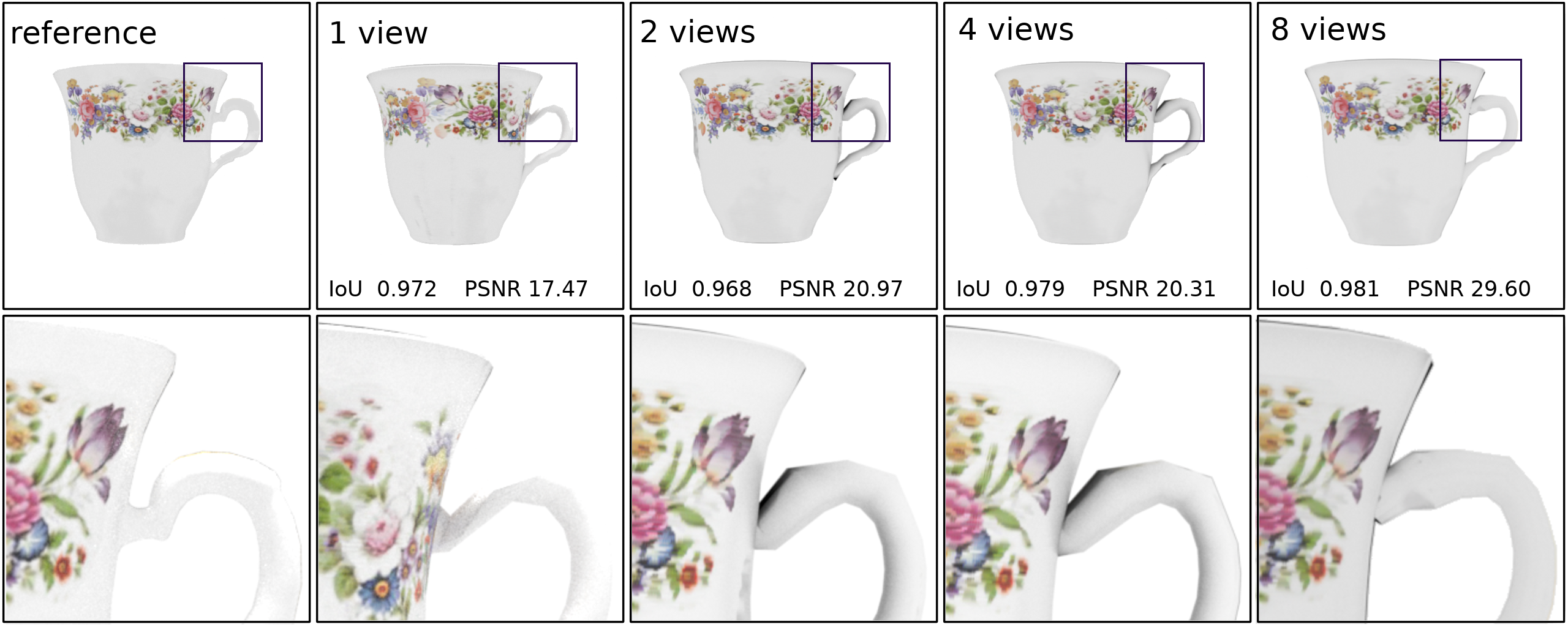}
	\caption{\label{fig_res_5}
		Results of 3D reconstruction with our approach and different viewpoints One viewpoint is enough for mesh reconstruction, and further increases do not make much improvement. However, more viewpoints allow to perform accurate texture reconstruction}
\end{figure}

We were mostly focused on a single-view reconstruction in this work, but our approach has support for multi-view reconstruction as well. However, we found out that adding new view usually don't lead to significant increase in quality, it can be useful for more complex objects or texture reconstruction. Figure~\ref{fig_res_5} shows how increasing number of views used for reconstruction affects the result. 

\subsection{Custom Differentiable Renderer}
\begin{figure}[H]
	\includegraphics[width=13 cm]{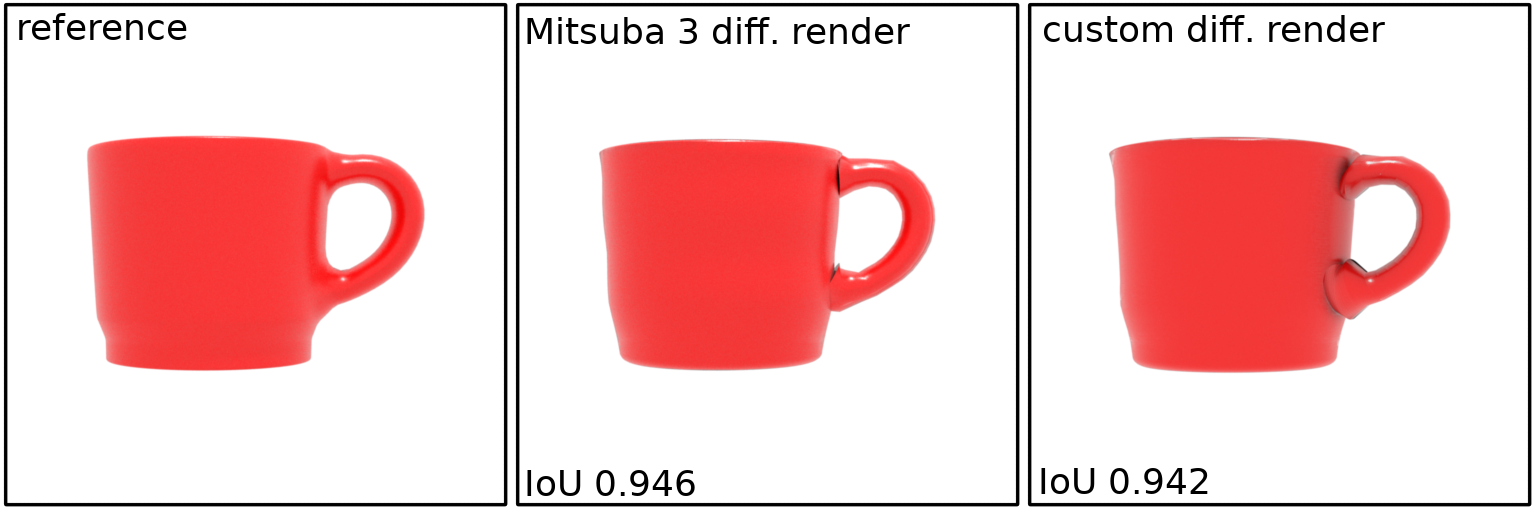}
	\caption{\label{fig_res_6}
		Reference image(right), reconstructed using Mitsuba 3(center) and our own differentiable renderer(right)}
\end{figure}

During the construction process most time is spent on differentiable rendering. The rendering is done is silhouette mode and it mean that we obtain vertex position gradients only from edge of the silhouette. It can be done by Mitsuba 3, but it there is no need in using such powerful tool for this simple task. Instead we created our own simple implementation of differentiable renderer that uses edge sampling\cite{li2018differentiable} to calculate the required derivatives. Figure~\ref{fig_res_6} shows that using it does not decrease quality of the reconstruction, but almost doubles it's speed. 

\section{Conclusion and Future Work}
In this work, we have presented a novel single-view 3D reconstruction approach that estimates the input parameters of a procedural generator to reconstruct the model. We implemented a few differentiable procedural generators and demonstrated how high quality results can be achieved using them. Also we proposed an alternative version of the same approach that do not rely on differentiability of the generator and can achieve decent quality using already existing non-differentiable procedural generators. We have implemented an efficient strategy for finding the optimal parameter sets for both versions of proposed approach. Our methods demonstrates better results compared to existing approaches and produces meshes with fewer artifacts. However, it has some limitations, primarily that it only works well on a class of objects that the underlying procedural generator can create. The differentiable procedural generators used in this work were created from scratch and are limited in their abilities. As future research, we plan to generalize our approach to a wider class of objects. It requires either integrating existing generators into our reconstruction pipeline or using generative neural networks.

\section{Acknowledgements}
The work was sponsored by the non-profit Foundation for the Development of Science and Education "Intellect".

\section{Discussion}

Authors should discuss the results and how they can be interpreted from the perspective of previous studies and of the working hypotheses. The findings and their implications should be discussed in the broadest context possible. Future research directions may also be highlighted.

\section{Contribution}

This work is the logical conclusion of our previous conference papers \cite{garifullin2022fitting} and \cite{garifullin2023diff} in which we combine the results for non-differentiable and differentiable procedural modeling. Our main contribution is that we propose a method to join differentiable rendering and inverse procedural modeling which gives several advantages: 

\begin{enumerate}
\item In comparison to existing general purpose 3D reconstruction approaches our method is more precise and artifact-free when a small number of input images are available. 
\begin{itemize}
	\item However, it is less general because it requires to implement a procedural generator for target class of 3D objects.
\end{itemize}

\item In comparison to existing inverse procedural modeling implementations our approach is more general: we have applied it for three different cases of inverse procedural modeling: trees, dishes and buildings. 
\begin{itemize}
	\item These generators are quite different from each other, and in each specific case 90\% changes is dedicated to loss function.
\end{itemize}   

\item By using differentiable rendering we propose a general and precise control method for inverse procedural modeling driven by input image. 
\begin{itemize}
	\item However, our method can work even if the gradient is not available. In this case only forward rendering is used and genetic algorithm does all the work which results to approximate solution of inverse modeling, driven by input image. 
\end{itemize}
	
\end{enumerate}

\begin{adjustwidth}{-\extralength}{0cm}

\reftitle{References}

\bibliography{bibliography.bib}

\end{adjustwidth}
\end{document}